\documentclass[aps,prfluids,showpacs,notitlepage,superscriptaddress,longbibliography]{revtex4-2}
\usepackage{epsfig,graphics,amssymb,amsmath,subeqnarray,color,bm,bbm}
\usepackage[toc,page]{appendix}
\usepackage{mathtools}
\usepackage{xcolor}
\usepackage{float}
\usepackage{graphicx}
\usepackage[colorlinks=true,allcolors=blue]{hyperref}

\usepackage{upgreek}
\usepackage{bm}

\usepackage{mathrsfs}

\begin{document}
\title{A comprehensive Darcy-type law for viscoplastic fluids: II.~Rheology \& topology}
\author{Emad Chaparian}
\email{emad.chaparian@strath.ac.uk}
\affiliation{James Weir Fluid Laboratory, Department of Mechanical \& Aerospace Engineering, University of Strathclyde, Glasgow, United Kingdom}
\date{\today}

%%%%%%%%%%%%%%%%%%%%%%%%%%%%%%%%%%%%
%%%%%%%%%%%%%%%%%%%%%%%%%%%%%%%%%%%%
%%%%%%%%%%%%%%%%%%%%%%%%%%%%%%%%%%%%
%%%%%%%%%%%%%%%%%%%%%%%%%%%%%%%%%%%%
\begin{abstract}
We extend our recently proposed framework (Chaparian, Phys.~Rev.~Fluids 10(9) 093301, 2025) for deriving a Darcy-type law governing viscoplastic flows through porous media to incorporate more applied aspects. In particular, the present work considers a more realistic rheological model (i.e.~Herschel-Bulkley, describing the shear-thinning nature of practical yield-stress fluids) along with a wider range of porous media topologies. In our earlier work, the problem was addressed by decomposing the full Bingham number spectrum (representing the ratio of the yield stress of the fluid to the characteristic viscous stress) into three main regions: (i) low Bingham numbers (weak yield stress limit) corresponding to Newtonian flow, (ii) high Bingham numbers (strong yield stress limit) representing yield limit/plastic flow, and (iii) intermediate Bingham numbers (transition regime). By deriving theoretical models for the two asymptotic limits of the spectrum and combining them, we obtained a Darcy-type law applicable across the entire range of Bingham numbers. In contrast to our original work, where the weak yield stress limit reduces to a Newtonian flow, here, this limit instead follows a power-law asymptote that captures the shear-thinning dominated behaviour of Herschel-Bulkley fluids. In the present study, we derive a scaling to address this limit. The framework is further generalised to incorporate a broader spectrum of porous media topologies, enabling a systematic assessment of how pore geometry influences the resulting macroscopic flow law. The proposed framework provides a unified theoretical basis for predicting yield-stress fluid transport through complex porous media and establishes a pathway towards finding macroscopic models applicable to a wide range of natural systems and industrial processes.
\end{abstract}
\maketitle
%%%%%%%%%%%%%%%%%%%%%%%%%%%%%%%%%%%%
%%%%%%%%%%%%%%%%%%%%%%%%%%%%%%%%%%%%
%%%%%%%%%%%%%%%%%%%%%%%%%%%%%%%%%%%%
%%%%%%%%%%%%%%%%%%%%%%%%%%%%%%%%%%%%

\section{Introduction}

Complex fluid flow through porous media is ubiquitous in many industrial processes \cite{frigaard2017bingham,zhou2018filtration,skauge2018polymer}, biomedical applications \cite{keating2003minimal} and geosciences \cite{walsh2008magma,gustafson1996prediction}. Due to the complex rheological behaviour exhibited by many practical fluids \cite{boger2013rheology,coussot2005rheometry}, macro-scale models for flows in porous media are typically semi-empirical. However, as reviewed in our previous work \cite{chaparian2025darcy1}, a few theoretical studies have proposed a more canonical model in which the pressure gradient is expressed as the sum of two terms associated with the characteristic viscous stress and the yield stress \cite{alfariss1987flow,talon2013determination,shahsavari2016mobility,bauer2019experimental,castaneda2023variational}. These two terms arise naturally from the physical scaling of the problem.

We rather approached this problem from a more theoretical perspective by splitting the entire problem into three main regimes based on the Bingham number (representing the ratio of the yield stress of the fluid to the characteristic viscous stress): (i) low Bingham number regime corresponding to the Newtonian flow, (ii) high Bingham number regime corresponding to the yield/plastic limit, and (iii) the intermediate Bingham numbers regime. By deriving models for the two asymptotic limits of the spectrum and combining them, we proposed a Darcy-type law applicable to the entire range of Bingham numbers \cite{chaparian2025darcy1}. In particular, we derived a universal scaling for the yield/plastic limit depends solely on the porosity of the porous medium: the non-dimensional critical pressure gradient required to initiate the yield-stress fluid flow inside a porous medium is scaled with the ratio of the obstructed space to the void space of the medium \cite{chaparian2024yielding}. For the low Bingham number limit--- Newtonian limit, we critically reviewed the Darcy laws proposed in the literature and identified the model that best represents the Newtonian regime through comparison with our extensive set of numerical simulations \cite{chaparian2025darcy1}.

The present study, which is a sequel to our original work \cite{chaparian2025darcy1}, extends this framework to include more realistic behaviours of yield-stress fluid flows in porous media. Specifically, we incorporate the shear-thinning rheology of practical yield-stress fluids and also investigate how different porous media topologies influence the macro-scale responses of these systems.

The proposed framework is founded on accurate description of the two asymptotic limits of the Bingham number spectrum. As demonstrated previously in a variety of yield-stress flow problems (e.g.~\cite{chaparian2017yield,frigaard2019background,chaparian2019porous,iglesias2020computing,chaparian2021sliding,frigaard2026u0}), the yield/plastic limit is independent of the viscous behaviour of the fluid. Consequently, ``simple'' yield-stress fluids exhibit the same critical pressure gradient for flow initiation, irrespective of their viscous behaviour. We show this feature both theoretically and computationally in the present study, implying that the scaling established in \cite{chaparian2024yielding} remains directly applicable in the present work. The other asymptotic limit, however, requires a different treatment. The weak yield stress limit was reduced to a Newtonian flow in our original work where the Bingham model was considered. Here, by adopting the Herschel-Bulkley constitutive equation to account for the shear-thinning behaviour, the corresponding asymptotic limit is analogous to power-law fluid flows. We therefore derive a new theoretical model for the low Bingham number regime that captures the dominant shear-thinning behaviour.

Shear-thinning fluid flow in porous media has been studied extensively since the last century in the context of enhanced oil recovery, e.g.~\cite{ikoku1979transient,odeh1979flow}. A nice review of these studies could be found in \cite{shah1995aspects}. More recently, related problems have also been investigated in fibrous media context, e.g.~\cite{shahsavari2015mobility}. Here, we tailor a shear-thinning power-law model for the low Bingham number limit based on a bound derived recently by Casta{\~n}eda \cite{castaneda2023variational} using a homogenisation approach.

Having addressed the two asymptotics, we propose our general Darcy-type law by combining these two models. To validate the proposed law, we perform an exhaustive series of Direct Numerical Simulations (DNS) with augmented Lagrangian method coupled with anisotropic adaptive mesh at the pore scale for a wide range of conditions (i.e.~various Bingham numbers, shear-thinning indices, porosities, topologies, etc.)

The outline of the paper is as follows. The next section \ref{sec:problem} sets the problem and introduces the generation procedure of random porous media and computational details. Section \ref{sec:plastic} revisits the yield/plastic limit, while the power-law limit (i.e.~low Bingham number regime) is addressed in section \ref{sec:powerlaw}. The resulting asymptotic models are combined in section \ref{sec:darcy} to derive the general Darcy-type law. Finally, the study is concluded in section \ref{sec:conclusion}.

\section{Problem set-up}\label{sec:problem}

\subsection{Mathematical formulation}

In this study, we consider incompressible 2D Stokes flows. For the rheological behaviour of the fluid, the Herschel-Bulkley constitutive equation is used,

\begin{equation}\label{eq:const}
	\left\{
	\begin{array}{ll}
		\hat{\boldsymbol{\uptau}} = \left( \hat{K} \Vert \hat{\dot{\boldsymbol{\upgamma}}} \Vert ^{n-1} + \displaystyle{\frac{\hat{\tau}_y}{\Vert \hat{\dot{\boldsymbol{\upgamma}}} \Vert}} \right) \hat{\dot{\boldsymbol{\upgamma}}} & ~~\mbox{iff}\quad \Vert \hat{\boldsymbol{\uptau}} \Vert > \hat{\tau}_y, \\[15pt]
		\hat{\dot{\boldsymbol{\upgamma}}} = \boldsymbol{0} & ~~\mbox{iff}\quad \Vert \hat{\boldsymbol{\uptau}} \Vert \leqslant \hat{\tau}_y,
	\end{array} \right.
\end{equation}
in which $\hat{K}$ is the consistency of the fluid and $n$ is the power-law index. In Eq.~(\ref{eq:const}), $\hat{\dot{\boldsymbol{\upgamma}}}$ is the rate-of-strain tensor or indeed $\boldsymbol{\nabla} \hat{\boldsymbol{u}} + \boldsymbol{\nabla} \hat{\boldsymbol{u}}^T$ where $\hat{\boldsymbol{u}}=(\hat{u},\hat{v})$ is the fluid velocity vector and $\hat{\boldsymbol{\uptau}}$ is the deviatoric stress tensor. The von Mises yielding criterion is considered for the fluid with the yield stress $\hat{\tau}_y$, i.e.~$\Vert \cdot \Vert$ is the second invariant of a tensor or $\Vert \boldsymbol{\Lambda} \Vert = \sqrt{(1/2) \boldsymbol{\Lambda} \boldsymbol{:} \boldsymbol{\Lambda}}$ when $\boldsymbol{\Lambda}$ is deviatoric. To make the equations non-dimensional, we use the following scalings,
\[
\left( x,y \right) = \frac{\left( \hat{x},\hat{y} \right)}{\hat{\ell}},~~\boldsymbol{u}=\left( u,v \right) = \frac{(\hat{u},\hat{v})}{\hat{U}}~~\&~~\left( p,\boldsymbol{\uptau} \right)=\frac{\left(\hat{p},\hat{\boldsymbol{\uptau}}\right)}{\hat{K} (\hat{U}/\hat{\ell})^n}.
\]
Indeed, the velocity vector is scaled with the interstitial mean velocity at the inlet ($\hat{U}$), and both the deviatoric stress tensor and pressure $\hat{p}$ with the characteristic viscous stress. The length scale $\hat{\ell}$ will be fixed later in this section. Therefore, the fluid flow in the void space (i.e.~in $\Omega \setminus \bar{X}$) is governed by the Cauchy equation and continuity,
\begin{equation}\label{eq:Stokes}
\boldsymbol{0} = - \boldsymbol{\nabla} p + \boldsymbol{\nabla} \boldsymbol{\cdot} \boldsymbol{\uptau} ~~\&~~\boldsymbol{\nabla} \boldsymbol{\cdot} \boldsymbol{u} = 0,~~~\text{in}~ \Omega \setminus \bar{X},
\end{equation}
and the non-dimensional form of the constitutive equation (\ref{eq:const}) reads,
\begin{equation}\label{eq:non-const}
  \left\{
    \begin{array}{ll}
      \boldsymbol{\uptau} = \left( \Vert \dot{\boldsymbol{\upgamma}} \Vert ^{n-1} + \displaystyle{\frac{B}{\Vert \dot{\boldsymbol{\upgamma}} \Vert}} \right) \dot{\boldsymbol{\upgamma}} & ~~\mbox{iff}\quad \Vert \boldsymbol{\uptau} \Vert > B, \\[15pt]
      \dot{\boldsymbol{\upgamma}} = \boldsymbol{0} & ~~\mbox{iff}\quad \Vert \boldsymbol{\uptau} \Vert \leqslant B,
  \end{array} \right. ~~~\text{in}~ \Omega \setminus \bar{X},
\end{equation}
where $B = \hat{\tau}_y / \left[ \hat{K} (\hat{U} / \hat{\ell} )^n \right] $ is the ratio of the yield stress of the fluid to the characteristic viscous stress known as the Bingham number.

Since the interstitial mean velocity is used as the velocity scale in this formulation, the non-dimensional flow rate is independent of the Bingham number and is simply equal to the interstitial inlet length $L_{\mathrm{inl}}$; see \cite{chaparian2021sliding,chaparian2024yielding,chaparian2025darcy1} for more details. Consequently, the macro-scale pressure gradient deriving the flow in a medium depends only on two free parameters: the Bingham number and the power-law index. Therefore in what follows, for brevity, we denote this dependence by two indices $(\Delta P/L)_{(B,n)}$. The macro-scale pressure gradient satisfies the energy balance equation:
\begin{align}\label{eq:energy}
\int_{\Omega \setminus \bar{X}}  ( \boldsymbol{\uptau} \boldsymbol{:} \boldsymbol{\dot{\upgamma}} ) ~\text{d} A & = \underbrace{ \int_{\Omega \setminus \bar{X}} ( \mu \dot{\boldsymbol{\upgamma}} \boldsymbol{:} \dot{\boldsymbol{\upgamma}} ) ~\text{d}A  }_{\text{viscous dissipation}} + \underbrace{ B \int_{\Omega \setminus \bar{X}} \Vert \dot{\boldsymbol{\upgamma}} \Vert ~\text{d}A }_{\text{plastic dissipation}} \nonumber \\[10pt]
& = \int_{\Omega \setminus \bar{X}} \Vert \dot{\boldsymbol{\upgamma}} \Vert^{n+1} ~\text{d}A + B \int_{\Omega \setminus \bar{X}} \Vert \dot{\boldsymbol{\upgamma}} \Vert ~\text{d}A \nonumber \\[10pt]
& = a(\boldsymbol{u},\boldsymbol{u}) + B j(\boldsymbol{u}) = \left( \frac{\Delta P}{L} \right)_{(B,n)} ~ \int_{\Omega \setminus \bar{X}} u ~\text{d}A = L(\boldsymbol{u}),
\end{align}
where $a(\boldsymbol{u},\boldsymbol{u})$ is the viscous dissipation, $B j(\boldsymbol{u})$ the plastic dissipation, and $L(\boldsymbol{u})$ the work done by the applied macro-scale pressure gradient to derive the fluid flux $\int_{\Omega \setminus \bar{X}} u ~\text{d}A$ through the porous medium.

\subsection{Porous media construction}
In the present study, the porous media are generated by randomly distributing mono-size obstacles, of either square or circle shapes, occupying the solid region $X$ in a computational domain $\Omega$ of size $\text{meas}(\Omega) = L \times L = 50 \times 50$. More details can be found in our previous works in the same context \cite{chaparian2021sliding,fraggedakis2021first,chaparian2024yielding,chaparian2025darcy1}. The solid ``volume'' fraction is denoted by $\phi=\text{meas}(X)/\text{meas}(\Omega)$ and as a result, the porosity of the medium (i.e.~the void fraction) is equal to $1-\phi$. The dimensionless area of each obstacle is fixed as $\pi$ which means that the length scale is $\hat{\ell}=\hat{L}_s/\sqrt{\pi}$ where $\hat{L}_s$ is the square edge length in the case of square obstacles and $\hat{\ell}$ is equal to the radius of obstacles in the case of circular obstacles.

\subsection{Methodology}

\begin{figure}
%\begin{center}
\includegraphics[width=0.8\textwidth]{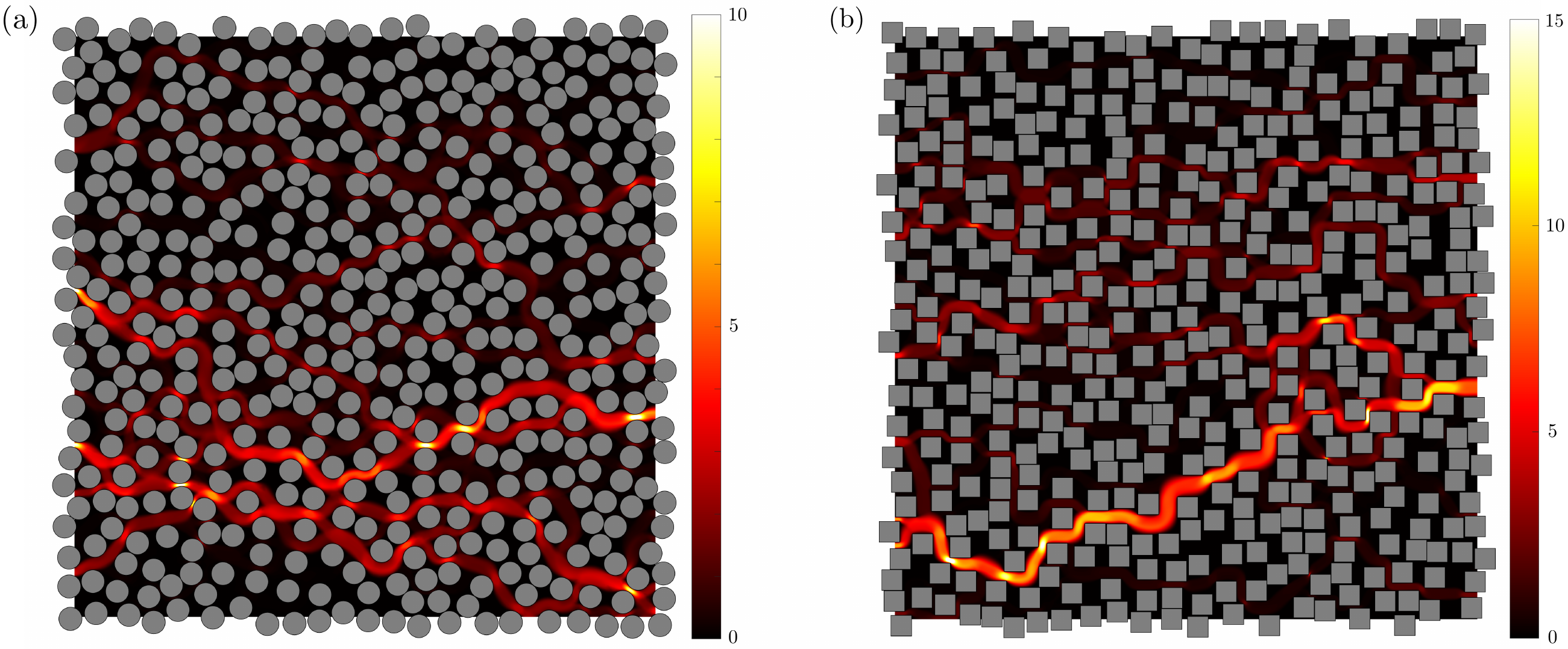}
\caption{Contour of velocity $\vert \boldsymbol{u} \vert$ for two sample simulations at $\phi=0.5$. In both panels, $B=10$ and $n=0.5$.}
\label{fig:B10}
%\end{center}
\end{figure}

Augmented Lagrangian method \cite{roquet2003adaptive,huilgol2005application,chaparian2019adaptive} is implemented in FreeFEM++ (an open-source finite element environment \cite{freefem}) and is coupled with anisotropic mesh adaptation at the pore scale; see Fig.~3 of \cite{chaparian2024yielding}. To highlight, the augmented Lagrangian method is capable of handling the non-differentiable viscoplastic models by relaxing the rate-of-strain tensor. We extensively validated this method for simulating yield-stress fluid flows in various physical settings over the past years; see for example \cite{chaparian2017yield,chaparian2019adaptive,chaparian2019porous,chaparian2022vane,medina2023rheo}. For the present problem, further details of the numerical implementation, including the boundary conditions and the imposing of the prescribed flow rate, are provided in \cite{chaparian2021sliding,fraggedakis2021first,parvar2024general,chaparian2024yielding,chaparian2025darcy1}. To illustrate the computational domain and the resulting flow fields, Fig.~\ref{fig:B10} presents two sample simulations at $B=10$ and $n=0.5$ for two different media with $\phi=0.5$.

\section{Yield/plastic limit: $B \rightarrow \infty$}\label{sec:plastic}

In this section, we revisit the high Bingham number limit which was fully addressed in our recent studies \cite{chaparian2024yielding,chaparian2025darcy1}. To highlight, in this limit ($B \rightarrow \infty$), the energy balance is dominated by plastic dissipation, therefore in the leading order, $B j(\boldsymbol{u}) \sim L(\boldsymbol{u})$ since the viscous dissipation is at least one order of magnitude smaller. Consequently, Eq.~(\ref{eq:energy}) predicts $(\Delta P / L)_{\infty} \sim B$ since both $j(\boldsymbol{u})$ and flux $\int_{\Omega \setminus \bar{X}} u ~\text{d}A$ reach their asymptotic values in this limit \cite{chaparian2019porous}. In other words, the pressure gradient threshold required to initiate flow is therefore independent of the power-law index, which is why only the subscript $_\infty$ is retained in the present notation. This is further confirmed by the computational data presented in \S \ref{sec:darcy}.

The dependence of $(\Delta P / L)_{\infty}$ on the topology and porosity of the porous media was addressed in our previous work \cite{chaparian2024yielding} by developing a model based on the energy equation and the preferred path of the fluid in this limit. The model was then fed with our statiscally-sound data generated by pore-network modelling \cite{fraggedakis2021first}, leading to the simple geometrical expression,
\begin{equation}\label{eq:plastic}
\frac{\left( \Delta \hat{P} / \hat{L} \right)_{\infty} }{\hat{\tau}_y / \hat{\ell}} = \pi \frac{\phi}{1-\phi} \Rightarrow \left( \frac{\Delta P}{L} \right)_{\infty} = \pi \frac{\phi}{1-\phi} B.
\end{equation}
As demonstrated in \cite{chaparian2024yielding}, although expression (\ref{eq:plastic}) is not a direct function of the topology of the medium, it remains highly accurate across a wide range of porous media topologies, including complex random media composed of bi-dispersed polygonal obstacles.

\section{Power-law limit: $B \rightarrow 0$}\label{sec:powerlaw}

In the present study, the low Bingham number limit differs fundamentally from the Newtonian limit considered in \cite{chaparian2025darcy1} because the fluid is Herschel-Bulkley. Instead, the asymptotic limit $B \rightarrow 0$ is analogous to power-law fluid flow. This has been acknowledged both in Al-Fariss \& Pinder's model \cite{alfariss1987flow} and Casta{\~n}eda's upper bound calculations \cite{castaneda2023variational}. As shown by Casta{\~n}eda, the upper bound of the macro-scale pressure gradient is,

%\begin{equation}\label{eq:castaneda_dim}
%\left( \frac{\Delta \hat{P}}{\hat{L}} \right)_{\left( \hat{\tau}_y,n \right)} \leqslant \hat{K} \frac{(1-\phi)^{\frac{1-n}{2}}}{\hat{\kappa}_{(0,1)}^{\frac{1+n}{2}}} ~\hat{U}_D^n + \sqrt{ \frac{1-\phi}{\hat{\kappa}_{(0,1)}} } ~\hat{\tau}_y
%\end{equation}

\begin{equation}\label{eq:castaneda}
\left( \frac{\Delta P}{L} \right)_{\left( B,n \right)} \leqslant \frac{(1-\phi)^{\frac{1-n}{2}}}{\kappa_{(0,1)}^{\frac{1+n}{2}}} ~\left( \frac{L_{inl}}{L} \right)^n + \sqrt{ \frac{1-\phi}{\kappa_{(0,1)}} } ~B,
\end{equation}
where $\kappa_{(0,1)}$ is the dimensionaless permeability of the medium for a Newtonian fluid (i.e.~when $B=0~\&~n=1$) which can be written as $ (L_{inl} / L) / (\Delta P/L)_{(0,1)}$. Hence, in the limit of low Bingham numbers ($B \rightarrow 0$),
\begin{equation}
\left( \frac{\Delta P}{L} \right)_{\left( 0,n \right)} \leqslant \left( \frac{\Delta P}{L} \right)^{\frac{1+n}{2}}_{(0,1)} \left( \frac{L_{inl}/L}{1-\phi} \right)^{\frac{n-1}{2}},
\end{equation}
or for a stochastic porous medium ($L_{inl}/L \equiv 1-\phi$), 
\begin{equation}\label{eq:castaneda_powerlaw}
\left( \frac{\Delta P}{L} \right)_{\left( 0,n \right)} \leqslant \left( \frac{\Delta P}{L} \right)^{\frac{1+n}{2}}_{(0,1)}.
\end{equation}
As mentioned, this is very close to Al-Fariss \& Pinder's model which in the limit of $B \rightarrow 0$ predicts,
\begin{equation}\label{eq:AlFarissPinder}
\left( \frac{\Delta P}{L} \right)_{\left( 0,n \right)} = \frac{1}{A(n)} \left( \frac{\Delta P}{L} \right)^{\frac{1+n}{2}}_{(0,1)},
\end{equation}
where $A(n)=4 \left( 3 + \frac{1}{n} \right) ^{-n} \left( \frac{3}{50} \right)^{\frac{1-n}{2}}$. Although these two models are derived using fundamentally different theoretical approaches, as discussed in our previous work \cite{chaparian2025darcy1} in detail, they exhibit several common features. Here, in both models, $(\Delta P/L) ^{(1+n)/2}_{(0,1)}$ is the scale of the power-law limit pressure gradient; only with a different pre-factor $1/A(n)$. 

\begin{figure}
%\begin{center}
\includegraphics[width=\textwidth]{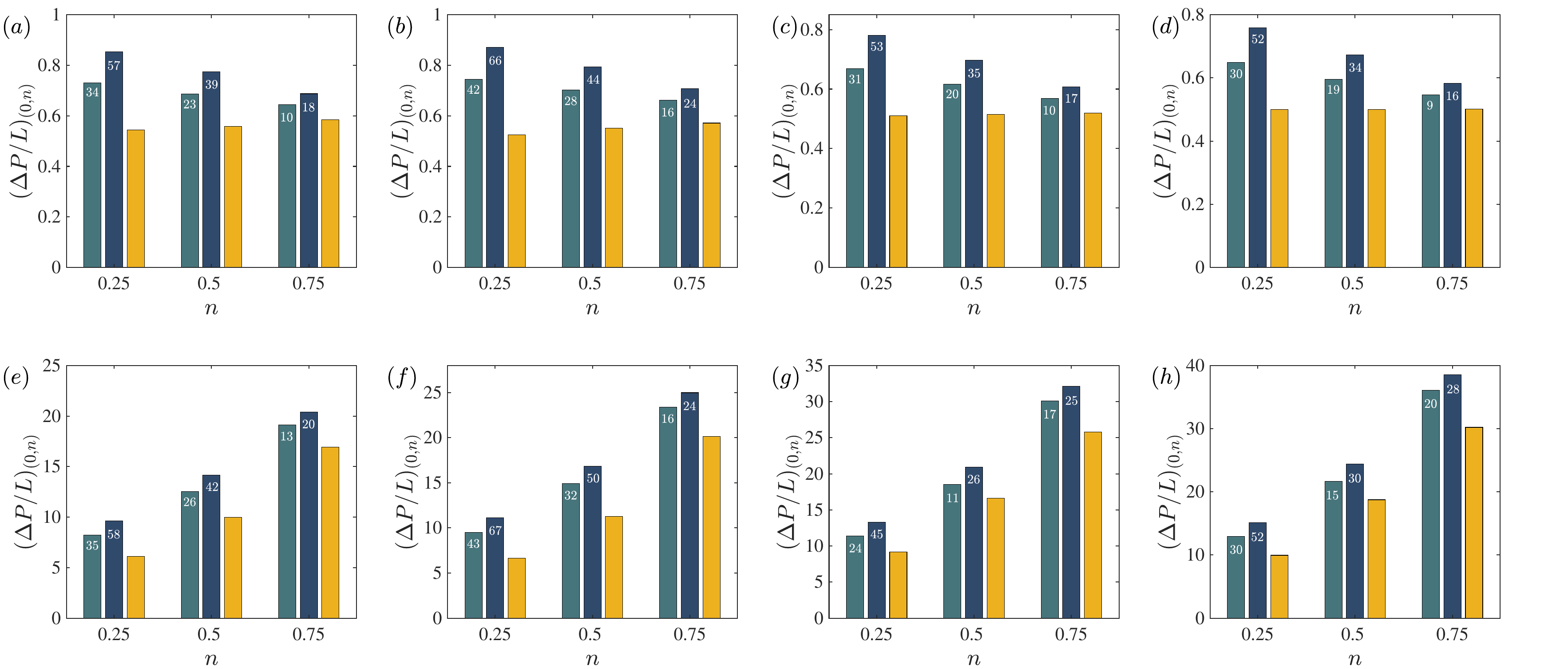}
\caption{Comparison of the predicted and computed macroscopic pressure gradients in the power-law limit ($B=0$) for 8 sample realisations for various power law indices ($n=0.25, 0.5 ~\&~ 0.75$). In panels (a-d) $\phi=0.1$ and in (e-h) $\phi=0.5$. Each panel represents one single realisation: (a,b,e,f) are circular obstacle cases and (c,d,g,h) are square obstacle cases. For each realisation, the left bar (green), the middle bar (blue) and the right bar (orange) correspond to the upper bound (\ref{eq:castaneda_powerlaw}), expression (\ref{eq:AlFarissPinder}), and the DNS result, respectively. The value on each bar indicates the uncertainty/error percentage relative to the DNS result, i.e.~the absolute difference between the theoretical prediction and the DNS result, normalised by the DNS value.
}
\label{fig:castaneda}
%\end{center}
\end{figure}

Fig.~\ref{fig:castaneda} compares these two expressions with the DNS results spanning three power law indices ($n=0.25, 0.5 ~\&~ 0.75$) for 8 sample realisations with relatively low and high solid ``volume'' fractions ($\phi=0.1~\&~0.5$), and both square and circular obstacle topologies.

\begin{figure}
%\begin{center}
\includegraphics[width=0.8\textwidth]{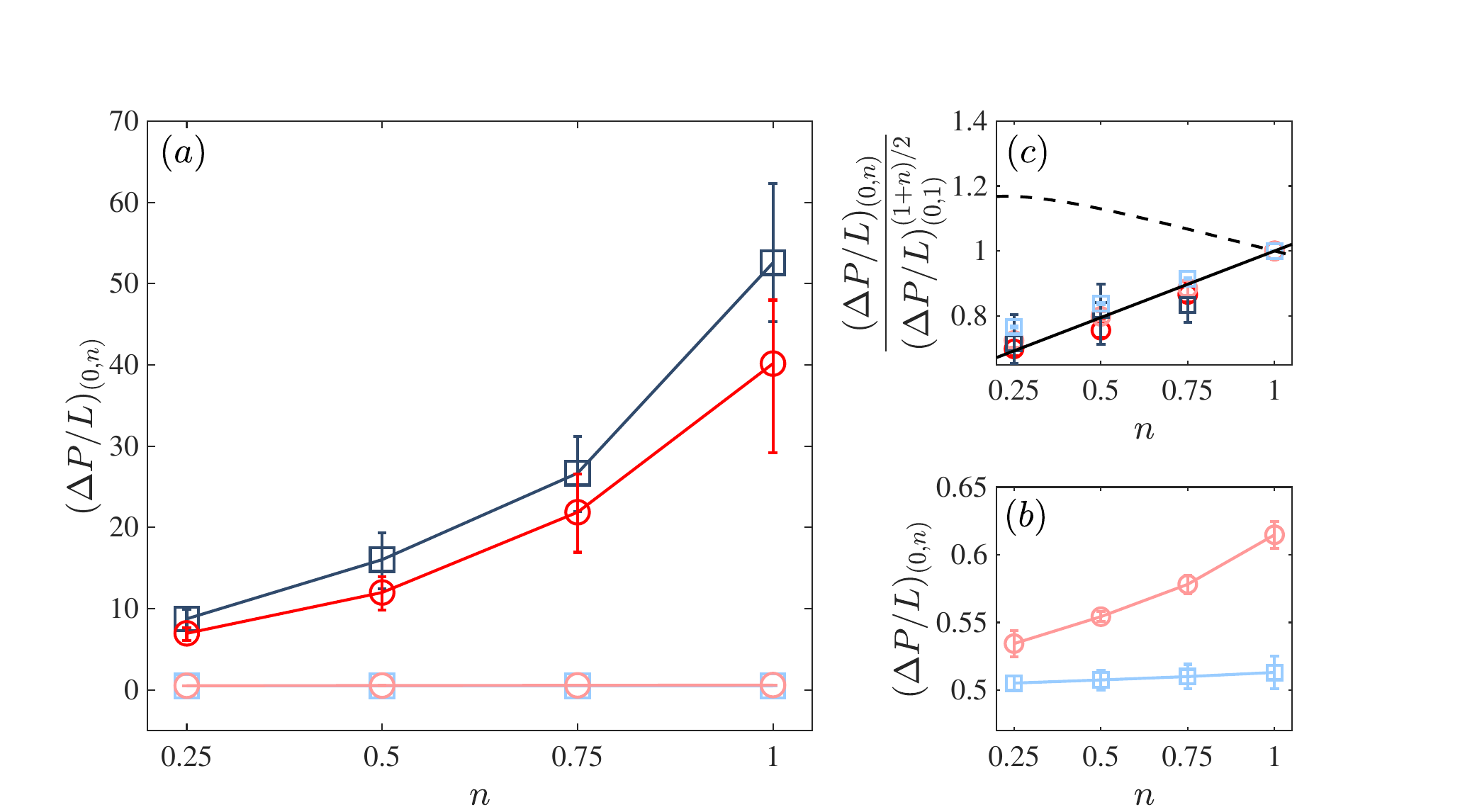}
\caption{Ensemble average (20 randomised realisations) of computed macro-scale pressure gradient at $B=0$ for various power-law indices $n=0.25, 0.5, 0.75 ~\&~1$. Uncertainty bars indicate the minimum and maximum values within each ensemble. Panel (a) plots $(\Delta P/L)_{(0,n)}$ for circular obstacle cases with red circular symbols and the case of square obstacles with blue square symbols. Dark and light colours correspond to $\phi=0.5$ and $\phi=0.1$, respectively. Panel (b) is an enlarged view of panel (a), highlighting the low pressure gradient values associated with $\phi=0.1$. Panel (c) shows the scaled pressure gradient: the DNS pressure gradient $\left( \Delta P/L \right)_{(0,n)}$ normalised by the scaling proposed in \cite{alfariss1987flow,castaneda2023variational}, i.e.~$\left( \Delta P/L \right)_{(0,1)}^{(1+n)/2}$. In panel (c), the dashed black line is $1/A(n)$ while the solid line shows the fitted ``master curve'', $g(n) = b~(n - 1)+1$, with $b=0.4102 \pm 0.013$ and RMSE equal to 0.0456.
}
\label{fig:Power-lawLimit}
%\end{center}
\end{figure}

As expected, the upper bound (\ref{eq:castaneda_powerlaw}) is consistently higher than the DNS data with a relatively moderate uncertainty. Expression (\ref{eq:AlFarissPinder}) yields an even larger overprediction of the computed values of $(\Delta P/L){(0,n)}$, since the function $A(n)$ is always less than unity for the practically relevant shear-thinning range $n \in [0,1)$. Nevertheless, Fig.~\ref{fig:castaneda} suggests that $(\Delta P/L)_{(0,1)}^{(n+1)/2}$ is a reasonable scaling for the macro-scale pressure gradient in the limit $B \rightarrow 0$. It should be mentioned that in Fig.~\ref{fig:castaneda} (and also later in Fig.~\ref{fig:Power-lawLimit}), the computed Newtonian pressure gradient $(\Delta P/L)_{(0,1)}$ from DNS is substituted in the two expressions (\ref{eq:castaneda_powerlaw}) and (\ref{eq:AlFarissPinder}).

To further examine this scaling, Fig.~\ref{fig:Power-lawLimit} presents the computational data for a large set of randomly generated realisations with two solid ``volume'' fractions and two types of obstacles, circular and square. Each data point represents the ensemble average of 20 different realisations with uncertainty bars indicating the minimum and maximum values within the ensemble. As expected, the macro-scale pressure gradient decreases with decreasing $n$ due to enhanced shear-thinning behaviour. Moreover, for diluter systems (or indeed higher porosities), $(\Delta P/L)_{0,n}$ is lower. More importantly, normalising the ensemble-averaged pressure gradient by the Newtonian value to the power $(n+1)/2$, as suggested by expressions (\ref{eq:castaneda_powerlaw}) \& (\ref{eq:AlFarissPinder}), collapses all data into a ``master curve'', irrespective of porosity, obstacle geometry, or power-law index; see Fig.~\ref{fig:Power-lawLimit}(c) and the solid black line in it. This suggests the following scaling,
\begin{equation}\label{eq:power_law}
\frac{(\Delta P/L)_{(0,n)}}{(\Delta P/L)_{(0,1)}^{(n+1)/2}} = g(n) = b~(n - 1)+1,
\end{equation}
where $b=0.4102$ is the fitted value. The function $g(n)$ therefore acts as a correction to the factor $1/A(n)$ in the Al-Fariss \& Pinder's model (shown by the dashed line in Fig.~\ref{fig:Power-lawLimit}(c)) and a pre-factor for Casta{\~n}eda's upper bound, thereby closing the gap between the theoretical predictions and the DNS data. This result provides the crucial low Bingham number asymptote required to formulate a general Darcy-type law for Herschel-Bulkley fluid flows through porous media.

\section{General Darcy-type law}\label{sec:darcy}
Having established the asymptotic models for the high Bingham number regime (yield/plastic limit) in \S \ref{sec:plastic} and the low Bingham number regime (power-law limit) in \S \ref{sec:powerlaw}, we now combine them to construct a general Darcy-type law. Following our framework proposed in \cite{chaparian2025darcy1} for Bingham fluids, the two asymptotic limits are linearly combined, giving:
\begin{align}\label{eq:Darcy_law}
\left( \frac{\Delta P}{L} \right)_{(B,n)} & = \left( \frac{\Delta P}{L} \right)_{(0,n)} + \left( \frac{\Delta P}{L} \right)_{\infty} \nonumber \\[10pt]
& = g(n) \left( \frac{\Delta P}{L} \right)_{(0,1)}^{\frac{1+n}{2}} + \left( \frac{\Delta P}{L} \right)_{\infty} \nonumber \\[10pt]
& = \left[ b~(n - 1)+1 \right] \left( \frac{\Delta P}{L} \right)_{(0,1)}^{\frac{1+n}{2}} + \pi \frac{\phi}{1-\phi} B,
\end{align}
where $\left( \Delta P/L \right)_{(0,1)}$ is the dimensionless macro-scale pressure gradient for a Newtonian flow through the porous medium. This is the only parameter which yet has to be determined theoretically and is the sole component of the proposed Darcy-type law (\ref{eq:Darcy_law}) through which the topology of the medium enters into the model.

\begin{figure}
%\begin{center}
\includegraphics[width=0.4\textwidth]{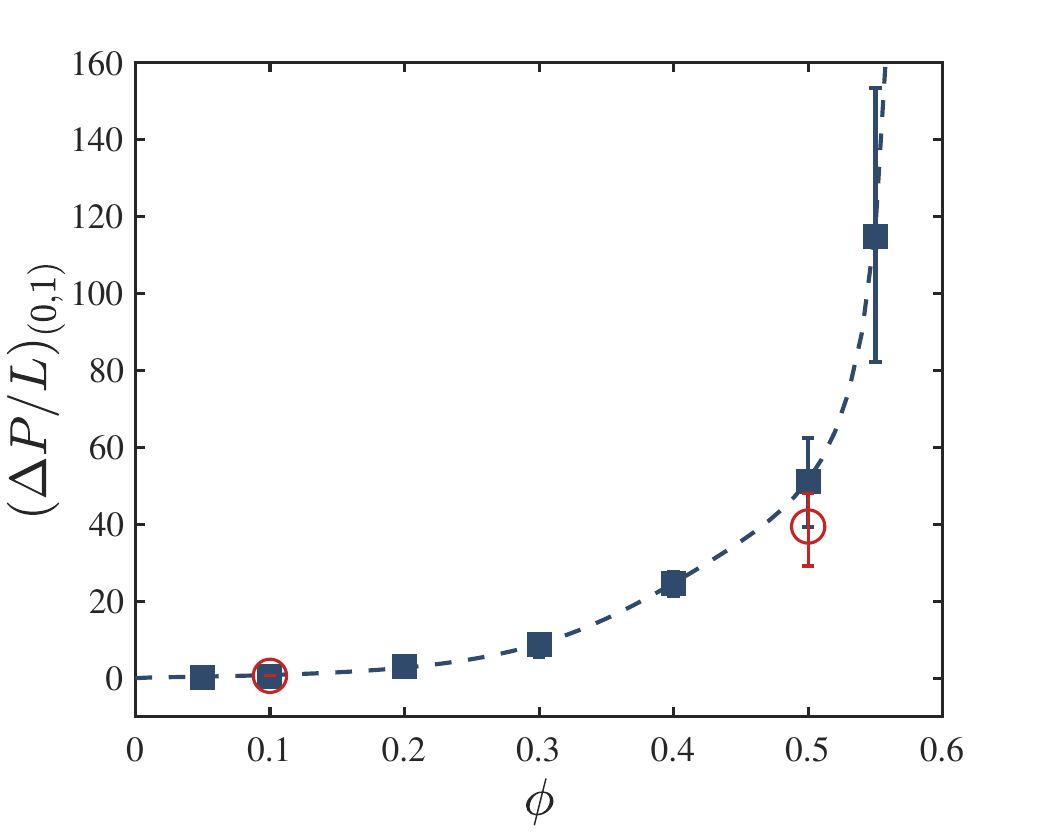}
\caption{Dimensionless macroscopic pressure gradient for Newtonian flow ($B=0$ and $n=1$) versus solid ``volume'' fraction. Solid blue square symbols are ensemble average of the DNS data obtained from 20 randomised mono-dispersed square samples at each $\phi$ while the open red symbols correspond to the circular obstacle cases. The dashed blue line is expression (\ref{eq:square}).}
\label{fig:topology}
%\end{center}
\end{figure}

For randomly generated porous media composed of mono-dispersed square obstacles, we previously adopted a Darcy law which is developed based on the Stokes drag on a 2D object (see \cite{chaparian2025darcy1} for details):
\begin{equation}\label{eq:square}
\left( \frac{\Delta P}{L} \right)_{(0,1)} = \frac{c}{-\log \phi + d + e ~\phi + f~ \phi^2 + O (\phi^3)}
\end{equation}
where $c=0.4091, d=-2.1954, e=4.1141 ~\&~ f=-2.1874$. Figure~\ref{fig:topology} compares the DNS data for $\left(\Delta P/L\right)_{(0,1)}$ obtained for randomly generated porous media composed of mono-dispersed square obstacles (blue solid square symbols) with the above expression (\ref{eq:square}) (dashed line). To assess the effect of the media topology on the Newtonian macro-scale pressure gradient, the corresponding DNS results for porous media composed of mono-dispersed circular obstacles are also shown for $\phi=0.1 ~\&~ 0.5$ (open red circles).

Although both geometries exhibit the same overall trend, a noticeable discrepancy emerges at lower porosities (i.e.~case $\phi=0.5$). Specifically, Newtonian flow through media composed of circular obstacles requires a smaller pressure gradient than the corresponding square obstacle case, consistent with the fact that the drag force on a cylinder is less than the drag force on a 2D object with a square cross section. In contrast, the two geometries' response are nearly identical for the diluter case (i.e.~$\phi=0.1$), indicating that topology becomes important only as the porosity decreases. Nevertheless, the same functional form given by expression (\ref{eq:square}) could potentially be applied to the circular obstacle cases by refitting the model coefficients against a sufficiently large dataset. Indeed, as discussed in our previous work \cite{chaparian2025darcy1}, the functional form of expression (\ref{eq:square}) was originally proposed for circular obstacles in \cite{drummond1984laminar}; see expression (14) of \cite{chaparian2025darcy1}.

\begin{figure}
%\begin{center}
\includegraphics[width=0.7\textwidth]{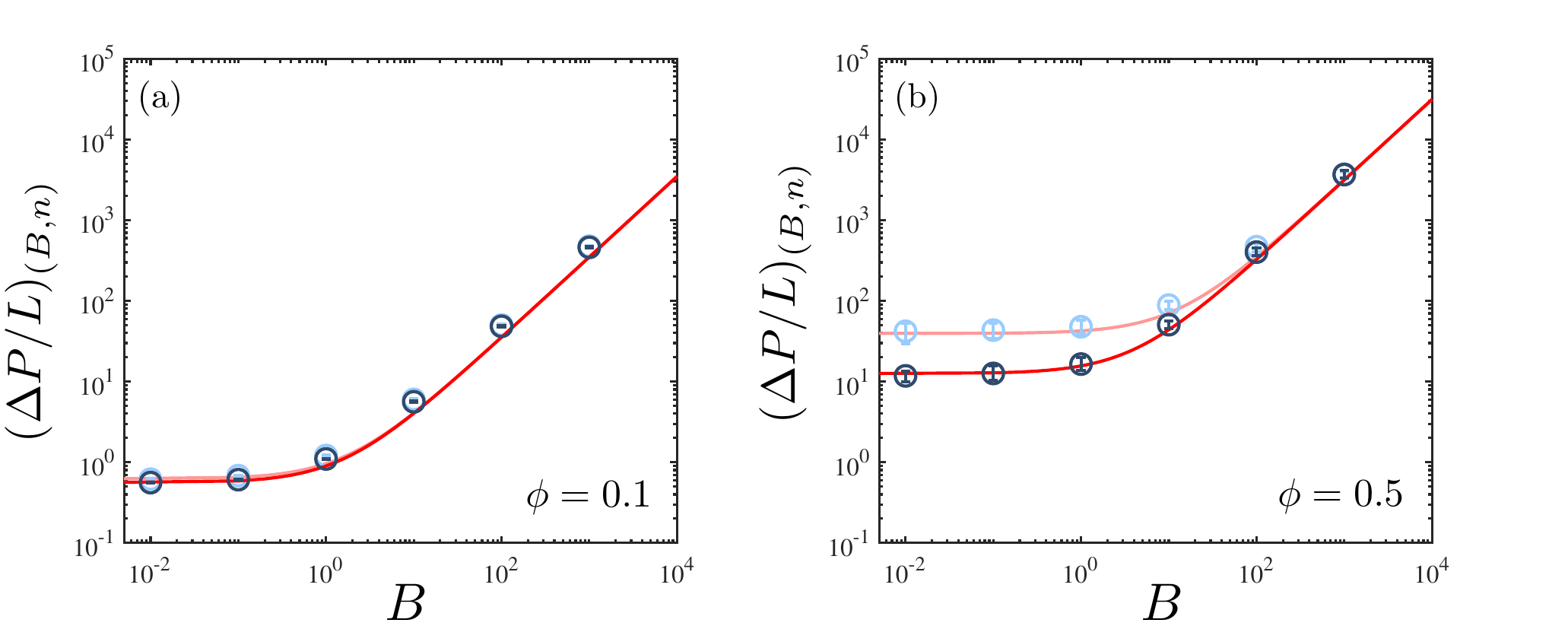}
\caption{Macro-scale pressure gradient for a Herschel-Bulkley fluid flow through randomised porous media made with circular obstacles: (a) $\phi=0.1$ and (b) $\phi=0.5$. Red curves are expression (\ref{eq:Darcy_law}), while symbols denote ensemble averaged DNS results with uncertainty bars. Dark and light colours correspond to $n=0.5$ and $n=1$, respectively.}
\label{fig:Darcy_law_circle}
%\end{center}
\end{figure}

Here, we instead substitute the Newtonian macroscopic pressure gradient for circular obstacle cases ($\phi=0.1~\&~0.5$) presented in Fig.~\ref{fig:topology} into our proposed Darcy-type law, i.e.~expression (\ref{eq:Darcy_law}). The resulting predictions are compared with DNS results in Fig.~\ref{fig:Darcy_law_circle}. Red curves represent the model predictions, while blue circular symbols denote ensemble averaged DNS results obtained from 20 independent realisations for each value of $\phi$. Dark and light colours correspond to $n=0.5$ and $n=1$ (Bingham fluid), respectively. Close agreement can be observed across the entire range of Bingham numbers. The comparison also shows that the influence of shear-thinning becomes more pronounced as the solid ``volume'' fraction increases. Moreover, in the high Bingham number regime, the computational data for both cases of $n=0.5$ and $n=1$ are nearly identical, showing the independence of the critical pressure gradient $(\Delta P/L)_{\infty}$ from the viscosity behaviour of ``simple'' yield-stress fluids discussed in \S \ref{sec:plastic}.

\begin{figure}
%\begin{center}
\includegraphics[width=0.9\textwidth]{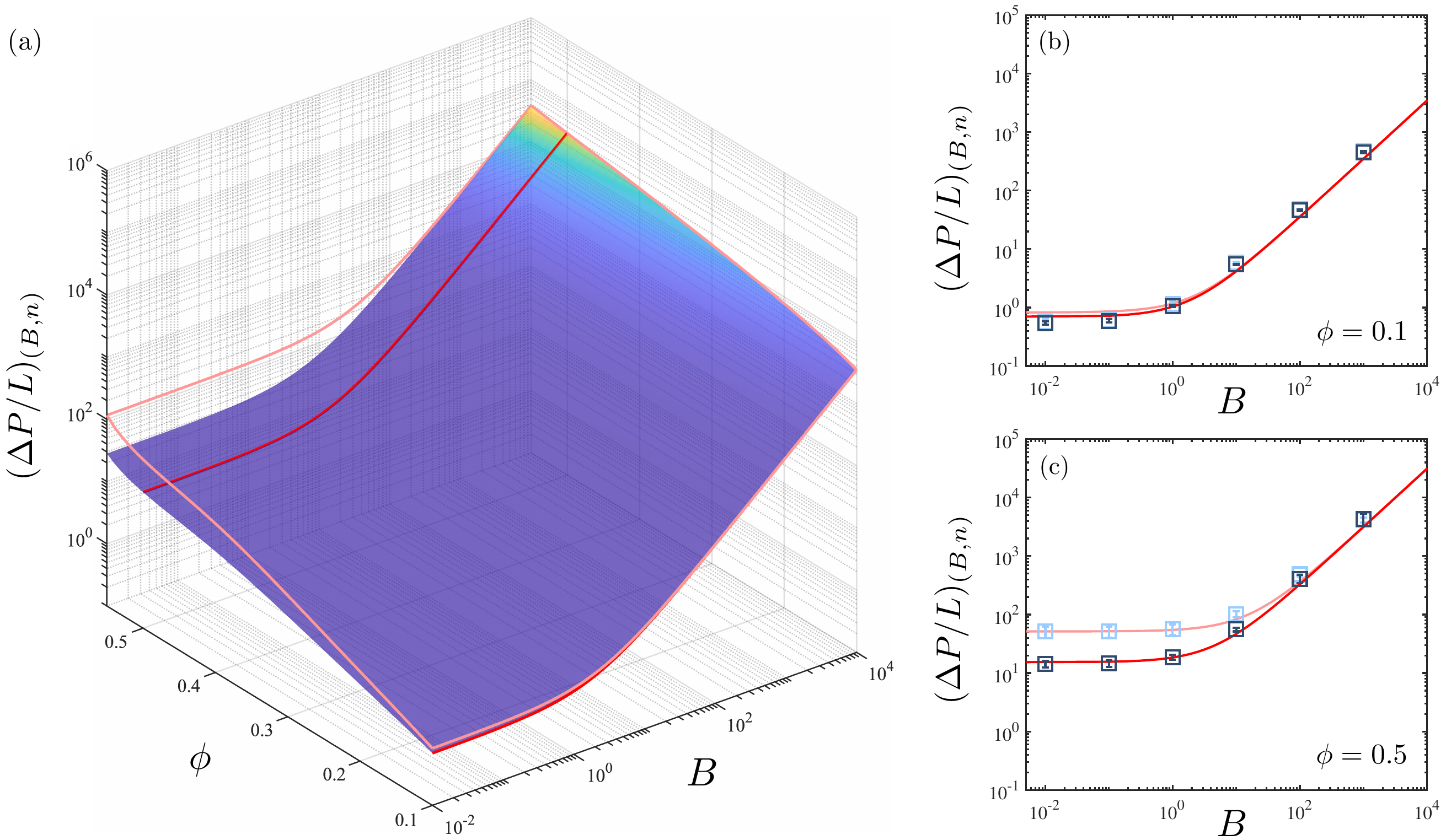}
\caption{Macro-scale pressure gradient for a Herschel-Bulkley fluid flow through randomised porous media made with square obstacles. The surface in panel (a) is expression (\ref{eq:darcy_square}) for the case $n=0.5$ with dark red curves indicating the cross sections at $\phi=0.1~\&~0.5$ which are also shown in panels (b) \& (c) with the same colour/style. The light red curves correspond to the Bingham fluid ($n=1$): in panel (a) forming the boundary of the surface in the case of a Bingham fluid, while in panels (b) \& (c) representing our proposed Darcy-type law for a Bingham fluid (i.e.~expression (30) in \cite{chaparian2025darcy1}). In panels (b) \& (c), symbols represent the ensemble average of the DNS data with uncertainty bars. Dark and light colours correspond to $n=0.5$ and $n=1$, respectively.}
\label{fig:Darcy_law_square}
%\end{center}
\end{figure}

Now, to examine the performance of our full Darcy-type law, we substitute $\left( \Delta P/L \right)_{(0,1)}$ of the square cases (expression (\ref{eq:square})) into the general form (\ref{eq:Darcy_law}), yielding,
\begin{equation}\label{eq:darcy_square}
\left( \frac{\Delta P}{L} \right)_{(B,n)} = \left[ b~(n - 1)+1 \right] \left( \frac{c}{-\log \phi + d + e ~\phi + f~ \phi^2 + O (\phi^3)} \right)^{\frac{1+n}{2}} + \pi \frac{\phi}{1-\phi} B.
\end{equation}
In the case of $n=1$ (i.e.~Bingham fluid), the above expression reduces to our Darcy-law proposed in \cite{chaparian2025darcy1}, i.e.~expression (30) in the reference. Expression (\ref{eq:darcy_square}) is shown as a surface in panel (a) of Fig.~\ref{fig:Darcy_law_square} for $n=0.5$. As expected, the pressure gradient increases monotonically with both the Bingham number and the solid ``volume'' fraction. The light red colour boundary marks the case of a Bingham fluid ($n=1$) where the limit $B \rightarrow 0$ reduces to a Newtonian flow instead. That is why the surface lies below this boundary due to the shear-thinning effects.

The comparison with the DNS data is made in panels (b) and (c) of Fig.~\ref{fig:Darcy_law_square} for $\phi=0.1$ and $\phi=0.5$ cases, respectively: the proposed law (\ref{eq:darcy_square}) is shown with the dark-red continuous line and the ensemble average DNS data (20 realisations for each $\phi$) with dark-blue square symbols. Excellent agreement is observed across the entire range of Bingham numbers, providing strong validity of the proposed Darcy-type law. For reference, the case of a Bingham fluid ($n=1$) is also shown in panels (b) and (c) using lighter colours.

\section{Summary \& final remarks}\label{sec:conclusion}

In this study, we extended our previously proposed Darcy-type law for Bingham fluid flows in porous media \cite{chaparian2025darcy1} to account for both the shear-thinning behaviour of yield-stress fluids and the influence of topology of the medium. The objective was to establish a unified macroscopic relation for the pressure gradient as a function of fluid rheological properties (i.e.~$B$ or the Bingham number---the ratio of the yield stress to the characteristic viscous stress---and the power-law index $n$), together with the porosity $(1-\phi)$ and the topology of the medium:
\[
\Delta P/L = f(B,n,\phi,\mathrm{topology}),
\]
which is shown by $(\Delta P/L)_{(B,n)}$ in short form for an individual medium where the solid ``volume'' fraction and the topology are fixed. For this aim, we followed our original framework proposed for a Bingham fluid and amended the asymptotic models to be able to capture the new effects. Indeed, in our original framework \cite{chaparian2025darcy1}, the entire range of Bingham numbers was decomposed into three regimes; two asymptotic low and high Bingham number regimes and an intermediate transition regime with moderate Bingham numbers.

The high Bingham number (yield/plastic) asymptote was retained from our previous work, where the critical pressure gradient (required to initiate the flow due to the yield stress of the fluid) was corroborated to depend only on the porosity of the porous medium; not the exact topology. Therefore, the geometrical scaling for the critical pressure gradient is,
\[
\left(\Delta P/L\right)_\infty=\pi\frac{\phi}{1-\phi}B.
\]
In addition, as shown in the present study, this asymptote is independent of the viscous behaviour of the fluid and therefore remains valid for any ``simple'' yield-stress fluid.

The principal extension introduced here concerns the low Bingham number limit. Unlike the Bingham model, for which the weak yield-stress limit reduces to a Newtonian flow, the Herschel-Bulkley fluid approaches a power-law asymptote. Building on earlier theoretical analyses \cite{alfariss1987flow,castaneda2023variational}, together with new exhaustive DNS results, we derived the scaling,
\[
(\Delta P/L)_{(0,n)} = g(n) ~(\Delta P/L)_{(0,1)}^{(1+n)/2},
\]
in which the correction function $g(n)=a(n - 1)+1$, found from the ``master curve'' fitted to the scaled computed pressure gradients, depends only on the power-law index. In other words, $g(n)$ is independent of both the porosity and the topology of the medium. Consequently, the influence of the medium microstructure/topology enters the proposed framework solely through the Newtonian pressure gradient $(\Delta P/L)_{(0,1)}$, thereby separating the effects of rheology from those of medium geometry. Indeed, two porous media with the same porosities can have different Newtonian permeabilities. This is more pronounced for media with lower porosities.

Combining the two asymptotic models yielded a general Darcy-type law applicable across the entire range of Bingham numbers. Extensive direct numerical simulations demonstrated excellent agreement with our proposed model over a wide range of Bingham numbers, power-law indices and porosities, providing strong validation of our theoretical framework.

The proposed Darcy-type law establishes a general theoretical framework for predicting yield-stress fluid flows through porous media while clearly separating the respective roles of fluid rheology, porosity of the medium and its microstructure/topology. Extending the present framework to fully three-dimensional and validating it against experimental measurements in our future studies facilitate its application to practical engineering and geophysical systems. Future studies should also broaden the framework to more general porous structures, including anisotropic media.

\appendix

\section{Comparison of expressions (\ref{eq:castaneda}) \& (\ref{eq:Darcy_law})}
\begin{figure}
%\begin{center}
\includegraphics[width=\textwidth]{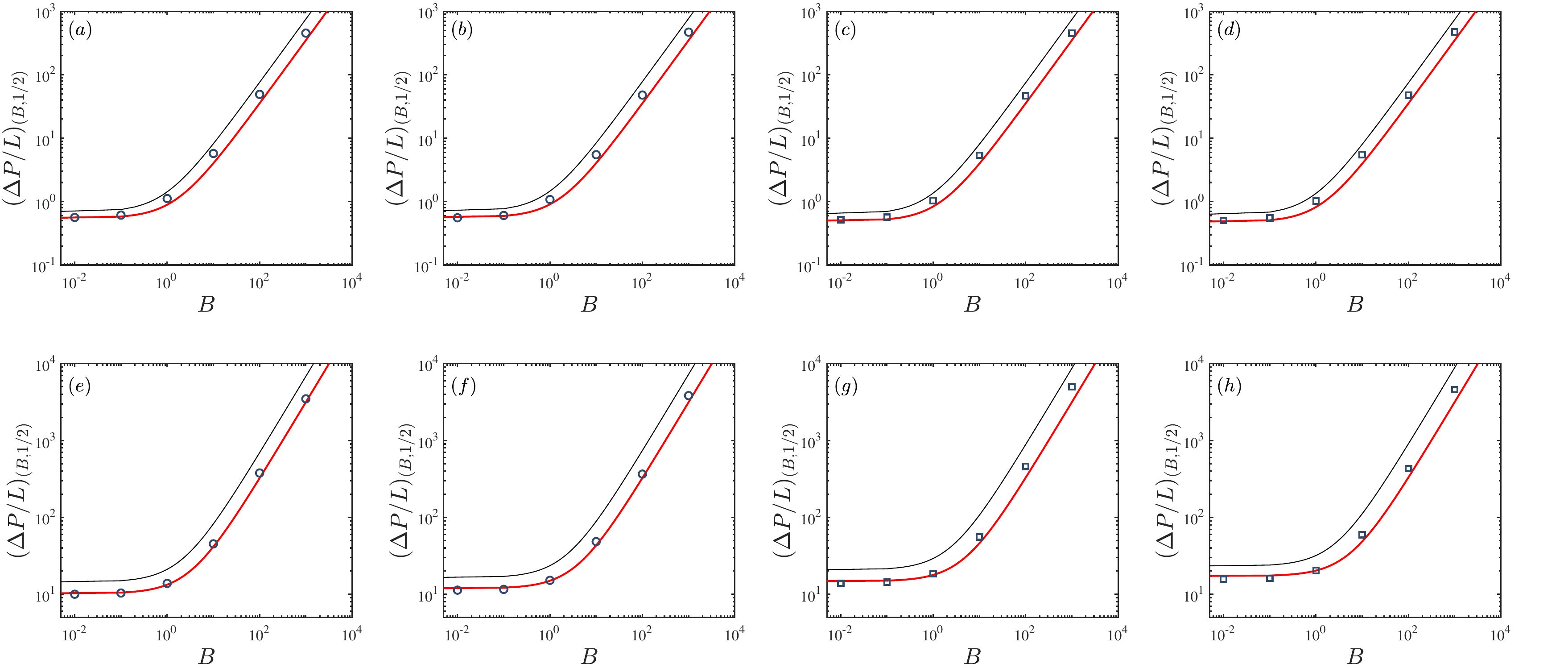}
\caption{Comparison of predicted and computed macroscopic pressure gradients for $n=0.5$ in the 8 sample realisations introduced in Fig.~(\ref{fig:castaneda}). Panels (a-d) correspond to $\phi=0.1$ and panels (e-h) to $\phi=0.5$. Each panel shows a single realisation: panels (a,b,e,f) represent circular obstacle media, and panels (c,d,g,h) represent square obstacle cases. In all panels, symbols denote DNS results, red lines represent expression (\ref{eq:Darcy_law}) and black lines expression (\ref{eq:castaneda}).}
\label{fig:appendix}
%\end{center}
\end{figure}

In this appendix, we compare Casta{\~n}eda's upper bound, expression (\ref{eq:castaneda}), and our proposed Darcy-type law (\ref{eq:Darcy_law}) with DNS data for the 8 sample realisations introduced in Fig.~\ref{fig:castaneda}. This is presented in Fig.~\ref{fig:appendix} with each panel corresponding to a single realisation: panels (a,b,e,f) show the circular obstacle cases and panels (c,d,g,h) the square obstacle cases. In all panels, symbols denote DNS results, red lines represent our proposed Darcy-type law, and black lines represent Casta{~n}eda's upper bound. Note that in all panels $n=0.5$. The proposed law agrees closely with the DNS results, and both lie below the upper bound throughout the cases shown.

\section*{Acknowledgment}

The author appreciates Sir David Anderson Bequest Award.

\section*{Data availability}

The data will be available from the authors upon request.

\bibliography{Viscoplastic}

\end{document}